# Experimental and theoretical studies on the photodegradation of 2-ethylhexyl 4-methoxycinnamate in the presence of reactive oxygen and chlorine species

Research Article

Alicja Gackowska[1], Maciej Przybyłek[2*], Waldemar Studziński[1], Jerzy Gaca[1]

[1]*Faculty of Chemical Technology and Engineering, University of Technology and Life Science, 85-326 Bydgoszcz, Poland*

[2]*Department of Physical Chemistry, Collegium Medicum, Nicolaus Copernicus University, 85-950 Bydgoszcz, Poland*

**Abstract:** 2-Ethylhexyl 4-methoxycinnamate (EHMC) is one of the most commonly used sunscreen ingredient. In this study we investigated photodegradation of EHMC in the presence of such common oxidizing and chlorinating systems as $H_2O_2$, $H_2O_2$/HCl, $H_2O_2$/UV, and $H_2O_2$/HCl/UV. Reaction products were detected by gas chromatography with a mass spectrometric detector (GC-MS). As a result of experimental studies chloro-substituted 4-methoxycinnamic acid (4-MCA), 4-methoxybenzaldehyde (4-MBA) and 4-methoxyphenol (4-MP) were identified. Experimental studies were enriched with DFT and MP2 calculations. We found that reactions of 4-MCA, 4-MBA and 4-MP with $Cl_2$ and HOCl were in all cases thermodynamically favorable. However, reactivity indices provide a better explanation of the formation of particular chloroorganic compounds. Generally, those isomeric forms of mono- and dichlorinated compounds which exhibits the highest hardness were identified. Nucleophilicity of the chloroorganic compounds precursors were examined by means of the Fukui function.

© Versita Sp. z o.o.

## 1. Introduction

Substances which absorb UV light and convert it into heat, fluorescence or phosphorescence, are often used as cosmetics ingredients and in photostabilization of polymeric materials and paints [1]. Because cinnamic acid derivatives convert absorbed light energy into heat with excellent efficiency [2,3] they are among the most popular UV sunscreen ingredients. Recently there has been a growing interest in the research of sunscreens [4-10].

There are many studies dealing with photodegradation of different compounds used as polymer additives [11], personal care products [10,12], dyes [13], and pharmaceuticals [14,15]. From the usability point of view, the UV absorbers should exhibit high photostability. Nevertheless there are studies which show that these compounds are relatively light-sensitive in environments of different polarity such as water [4,12], methanol [16], and petroleum jelly [17]. MacManus-Spencer *et al.* [12] found that 2-ethylhexyl 4-methoxycinnamate (EHMC) undergoes photolysis which results in formation of 4-methoxybenzaldehyde and 2-ethylhexyl alcohol. From the theoretical point of view, stability of organic compounds can be evaluated using thermochemical calculations. In many cases however the thermodynamic approach does not provide a complete description and hence reactivity indices should be taken into account.

* E-mail: m.przybylek@cm.umk.pl

**Table 1.** Molar proportions and concentrations of the reagents in the studied systems.

| Molar proportion | Concentration [M] | | |
|---|---|---|---|
| | EHMC | $H_2O_2$ | HCl |
| 1:10:10 | 0.005 | 0.05 | 0.05 |
| 1:5:10 | 0.005 | 0.025 | 0.05 |
| 1:10:5 | 0.005 | 0.05 | 0.025 |

It is well known that reactive oxygen and chlorine species (ROCS) are formed in many natural and industrial processes. Hydrogen peroxide and its mixtures with HCl are the sources of many various chlorinating and oxidizing agents such as $OH^\bullet$, $HO_2^\bullet$, $O_2$, $Cl^\bullet$, $Cl^-$, $Cl^+$, $Cl_2$, HOCl, $H_2OCl^+$, $ClO^-$ [18]. According to Nakajima et al. [19] chlorinated EHMC derivatives can be formed during EHMC photodegradation in the presence of HOCl. As we reported in our previous paper, chlorinating and oxidizing agents significantly enhance photodegradation of UV absorbers [20]. The main objective of this study has been to research EHMC photodegradation in the presence of common oxidizing and chlorinating model systems such as $H_2O_2$, $H_2O_2$/HCl, $H_2O_2$/UV and $H_2O_2$/HCl/UV [14,18,21-23]. In order to achieve this goal the experimental methods, such as GC/MS chromatography, spectrophotometric measurements as well as the theoretical tools (thermodynamic calculations, reactivity indices) were applied.

## 2. Experimental procedure

### 2.1. Materials and methods

All chemicals were used without further purification. Analytical standard of 2-ethylhexyl 4-methoxycinnamate (EHMC) (98%) was obtained from ACROS Organics (USA) and was kept in lightproof container at 4°C. Other chemicals used in the studies, methanol MeOH (96%), ethyl acetate, $H_2O_2$ (30%) and HCl (36%) were purchased from POCH S.A. (Poland).

### 2.2. Reaction conditions

First, methanolic solutions of EHMC, $H_2O_2$ and HCl were prepared. These solutions were then mixed in molar proportions according to Table 1. 900 mL of each mixture was subjected to the UV light using a photoreactor equipped with a medium pressure mercury lamp (Heraeus, TQ 150W), which emits radiation in the 200-600 nm range. The mixtures in the photoreactor were stirred using a magnetic stirrer (200 rpm). The temperature of the system was kept at 20°C.

### 2.3. Analytical methods

Changes in EHMC concentration were monitored using Agilent 8452A spectrophotometer. The absorbance changes of the EHMC solution were measured at $\lambda_{max}$ = 310 nm. Samples of 0.1 ml were taken from the reactor at different time intervals and diluted 60 times with methanol before measuring the spectra. All spectrophotometric measurements were performed in standard quartz cuvettes of 1 cm optical path. Methanol was used as a blank sample. Each photodegradation experiment was performed three times and the mean values of absorbance were used for kinetic analysis.

Qualitative analyses of reaction mixtures were performed according to the following procedure. After UV radiation, solutions (ca. 900 mL) were concentrated to a volume of 1 mL using rotary vacuum evaporator (p=122 mbar, t=40°C). In the next step, 0.1 mL of concentrate was diluted with 1 mL of ethyl acetate. So prepared samples were analyzed by HEWLETT PACKARD 5890 gas chromatographer equipped with a mass spectrometric detector and the ZB-5MS column (0.25 mm × 30 m × 0.25 µm). The operating conditions were as follows: 100°C-10°C min$^{-1}$-200°C/2 min-10°C min$^{-1}$-300°C 2 min$^{-1}$. The volume of the sample was 1 µL. Helium was used as a carrier gas. The schematic diagram of the experimental procedures is illustrated in Fig. 1.

## 3. Calculation details

All geometry optimizations and thermodynamic analysis, were performed applying the Becke three-parameter (B3) hybrid method [24-26] with the Lee–Yang–Parr (LYP) functional [27] and 6-31+G(d,p) basis set (B3LYP/6-31+G(d,p)). For open-shell systems, the spin-unrestricted calculations were applied (UB3LYP/6-31+G(d,p)). In order to get more accurate results, a large integration grid and tight convergence criteria were used. Vibrational analysis was performed for all compounds and no imaginary frequencies were found. Changes of thermodynamic functions, enthalpy (ΔH) and Gibbs free energy (ΔG), were calculated for standard conditions (T=298.15 K and P=101325 Pa). For each compound only the most stable conformation of a molecule was taken into account (global minima on the potential energy hypersurface). In order to include the solvent effect the self-consistent reaction field (SCRF) with the polarized continuum model (PCM) [22,23] and Bondi-type parameterization was used. All density functional theory (DFT) calculations presented here were performed using Gaussian 03 software [30]. Global reactivity indices, such as absolute electronegativity χ, hardness η [31], and electrophilicity ω [32] were calculated from the Eqs. 1-3:

$$\chi = -(E_{LUMO} + E_{HOMO})/2 \quad (1)$$

$$\eta = (E_{LUMO} - E_{HOMO})/2 \quad (2)$$

$$\omega = \chi^2/2\eta \quad (3)$$

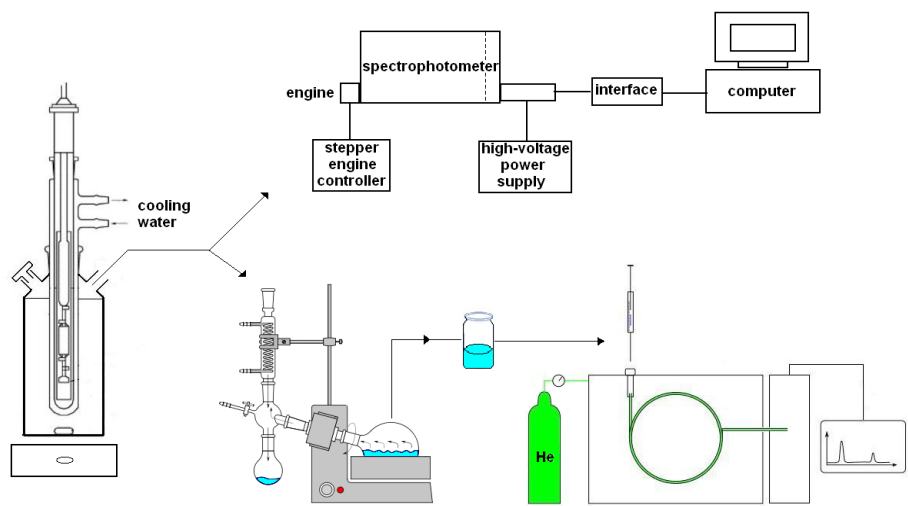

**Figure 1.** Schematic diagram of experimental procedures.

Energy of the lowest unoccupied molecular orbital, $E_{LUMO}$ and the highest occupied molecular orbital, $E_{HOMO}$ were calculated for the optimized at B3LYP/6-31+G(d,p) structures using the Møller-Plesset perturbation method [33-38] and 6-31+G(d,p) basis set (MP2/6-31+G(d,p)). In order to characterize aromaticity of the studied compounds, a harmonic oscillator model of aromaticity (HOMA) index [39,40] was calculated according to the following equation:

$$HOMA = 1 - (\alpha/N)\Sigma(R_{opt} - R_i)^2 \quad (4)$$

where $N$ is the number of bonds in the aromatic ring while α and $R_{opt}$ are constants depending on the type of bond (for C−C bond α is 257.7, and $R_{opt}$ is 1.388 Å). The condensed Fukui function $f_j^-$ was calculated from the Eq. 5 according to the procedure proposed by Yang and Mortier [41]:

$$f_j^- = q_j(N) - q_j(N-1) \quad (5)$$

where q is NBO charge population calculated at the UB3LYP/6-311++G(2d,2p) level while N denotes number of electrons in the optimized structure.

## 4. Results and discussion

### 4.1. Characterization of the reaction mixture

In order to examine the effect of oxidizing and chlorinating agents on EHMC photodegradation, the following model systems were prepared: EHMC; EHMC/$H_2O_2$; EHMC/$H_2O_2$/HCl. Spectrophotometric measurements showed that in non-irradiated solutions a decrease in absorbance at $\lambda_{max}$ = 310 nm was very small. In case of the EHMC/$H_2O_2$ system the absorbance diminished by about 10% after 28 days while the presence of HCl slightly accelerated the EHMC degradation.

After 24 hours of irradiation of the EHMC solution without HCl and $H_2O_2$ the absorbance was reduced by about 24% (Fig. 2a). Addition of $H_2O_2$ significantly influenced the rate of EHMC photodegradation. After 10 hours of irradiation the absorbance decreased by half. As one can see from Fig. 2a, the highest rate of EHMC conversion was achieved in case of the EHMC/$H_2O_2$/HCl/UV system. When the molar ratio of the EHMC/$H_2O_2$/HCl system was 1:10:10 photodegradation was considerably faster than in other cases (Fig. 2b). The GC/MS analysis showed that the main reactions occurring in the EHMC/$H_2O_2$/UV system were trans-cis photoisomerisation and photolysis, which confirmed the presence of 2-ethylhexyl alcohol (EHA) and Z-EHMC in the reaction system (Fig. 3). Noteworthy, this is consistent with the results reported in earlier studies [12,42,43].

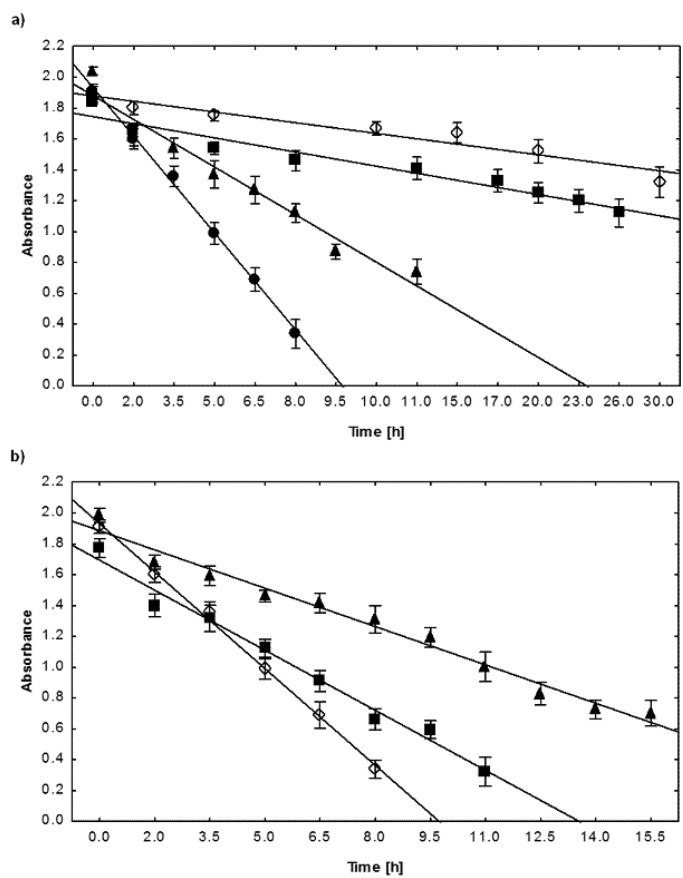

**Figure 2.** Changes in EHMC absorbance during UV irradiation: a) impact of the presence of $H_2O_2$ and HCl: —◊— EHMC, —■— EHMC/HCl [1:10], —▲— EHMC/$H_2O_2$ [1:10], —●— EHMC/$H_2O_2$/HCl [1:10:10], b) impact of molar ratio of EHMC/$H_2O_2$/HCl system: —◊— [1:10:10], —■— [1:10:5], —▲— [1:5:10].

**Table 2.** Mass spectral data of non-chlorinated compounds detected in all reaction mixtures and chloroorganic products detected in the EHMC/$H_2O_2$/HCl/UV system.

| Compound | Abbreviation | Main m/z peaks (% of base peak) |
|---|---|---|
| 2-ethylhexyl 4-methoxycinnamate | EHMC | 178 (100), 161 (53), 134 (17), 133 (19) |
| 2-ethylhexyl alcohol | EHA | 83 (18), 70 (20), 57 (100), 41 (40) |
| 4-methoxycinnamic acid | 4-MCA | 178 (100), 161 (37), 133 (22), 89 (18) |
| 4-methoxybenzaldehyde | 4-MBA | 135 (100), 107 (12), 92 (19), 77 (38) |
| 4-methoxyphenol | 4-MP | 124 (81), 109 (100), 81 (58), 53 (23) |
| 3-chloro-4-methoxycinnamic acid | 3-Cl-4-MCA | 212 (100), 176 (55), 132 (38), 89 (23) |
| 3,5-dichloro-4-methoxycinnamic acid | 3,5-diCl-4-MCA | 246 (100), 229 (46), 202 (32), 73 (21) |
| 3-chloro-4-methoxybenzaldehyde | 3-Cl-4-MBA | 169 (100), 126 (11), 99 (18), 63 (25) |
| 3,5-dichloro-4-methoxybenzaldehyde | 3,5-diCl-4-MBA | 203 (100), 160 (60), 209 (10), 65 (20) |
| 3-chloro-4-methoxyphenol | 3-Cl-4-MP | 158 (61), 143 (100), 107 (38), 55 (33) |
| 2,5-dichloro-4-methoxyphenol | 2,5-diCl-4-MP | 192 (56), 177 (100), 149 (33), 53 (41) |

The EHMC conversion significantly increased with the addition of HCl to the system (Fig. 2a). As one can see from Fig. 3, much more of photodegradation products were formed during EHMC UV irradiation in the presence of chlorinating agents than in case of the EHMC/$H_2O_2$/UV system. The following chloroorganic compounds, 3-chloro-4-methoxybenzaldehyde (3-Cl-4-MBA), 3,5-dichloro-4-methoxybenzaldehyde (3,5-diCl-4-MBA), 3-chloro-4-methoxyphenol (3-Cl-4-MP), 2,5-dichloro-4-methoxyphenol (2,5-diCl-4-MP), 3-chloro-4-methoxycinnamic acid (3-Cl-4-MCA), and 3,5-dichloro-4-methoxycinnamic acid (3,5-diCl-4-MCA) were identified in the reaction mixture (Table 2). Interestingly, we did not detect chloro-substituted EHMC derivatives. Based on the detected products in the EHMC/$H_2O_2$/UV and EHMC/$H_2O_2$/HCl/UV systems (Table 2), we proposed the following schematic pathways of EHMC photodegradation in the presence of ROCS (Fig. 4).

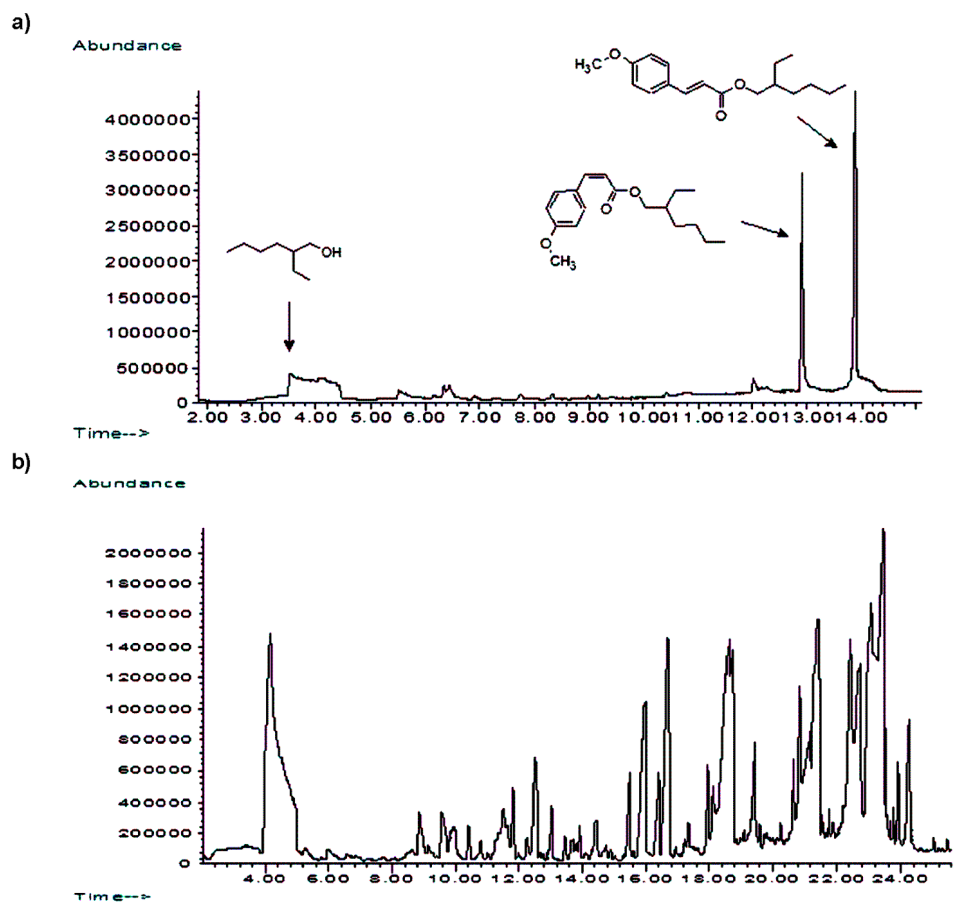

**Figure 3.** Gas chromatograms of the products of EHMC conversion in: a) EHMC/$H_2O_2$/UV [1:10] system and in b) EHMC/$H_2O_2$/HCl/UV system [1:10:10].

### 4.2. Thermodynamic insight into formation of detected products

Because the reaction proceeds very slowly without UV irradiation, the EHMC degradation is probably mainly photochemical and radical in nature. Photodissociation reactions of EHMC and $H_2O_2$ can be characterized by the bond dissociation enthalpy (BDE) defined as the enthalpy of reactions presented in Figs. 5a and 5b. According to the B3LYP/6-31+G(d,p) thermochemical calculations, the values of $H_2O_2$ and EHMC BDE are respectively 69.70 kcal mol$^{-1}$ and 44.20 kcal mol$^{-1}$, corresponding to 410.33 nm and 646.66 nm when expressed in wavelength. This shows that photodegradation reactions can occur under irradiation provided by the light source used in the experiment. It is worth noting that the C–O bond cleavage in the alkoxy moiety (BDE=117.60 kcal mol$^{-1}$) is less favorable than in case of the C–O bond cleavage between the carbonyl carbon and the alkoxy oxygen presented in Fig. 5a. 4-Methoxycinnamoyl radical formed *via* EHMC photolysis is probably further oxidized to 4-MCA, 4-MBA, and 4-MP. Although hydroxyl radicals are very short-lived species they are regarded as oxidizing agents in many photoinduced reactions between organic compounds and $H_2O_2$ [44-46]. As we found, $\Delta H$ and $\Delta G$ values of the reaction between 4-methoxycinnamoyl and •OH are -101.10 kcal mol$^{-1}$ and -89.59 kcal mol$^{-1}$, respectively. Highly favorable reaction with •OH is characteristic for antioxidants. It is worth mentioning that the growth of EHMC antioxidant activity during UV irradiation was experimentally proved by Dobashi *et al.* [1].

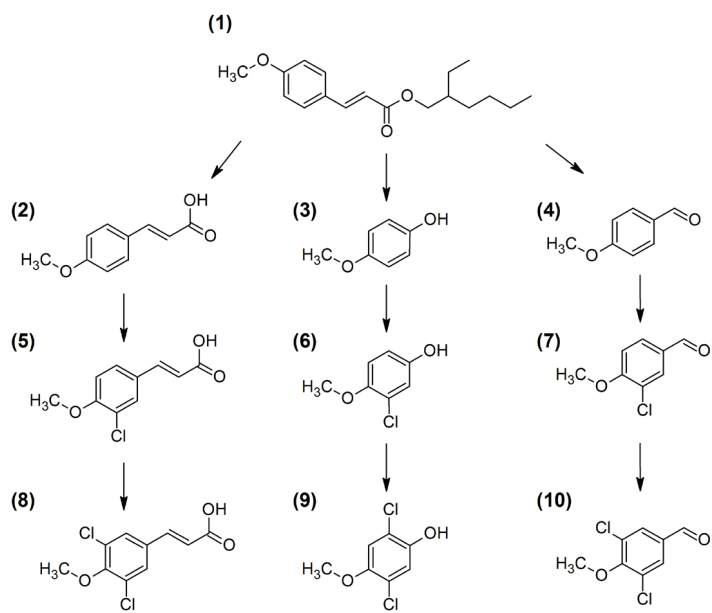

**Figure 4.** Proposed mechanistic pathways for EHMC photodegradation in presence of $H_2O_2$ and HCl; (1) E-EHMC, (2) 4-MCA, (3) 4-MP, (4) 4-MBA, (5) 3-Cl-4-MCA, (6) 3-Cl-4-MP, (7) 3-Cl-4-MBA, (8) 3,5-diCl-4-MCA, (9) 2,5-diCl-4-MP, (10) 3,5-diCl-4-MBA.

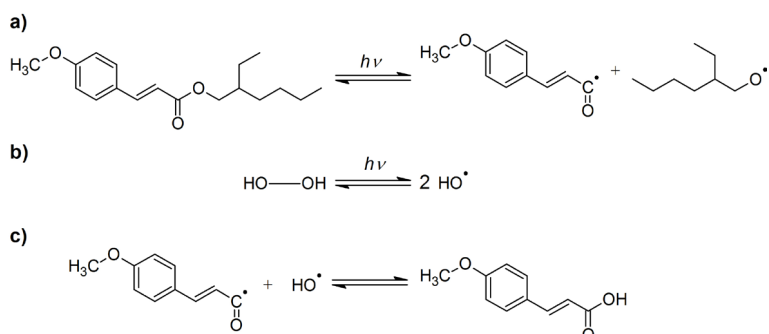

**Figure 5.** Radical reactions of EHMC and $H_2O_2$ occurring under the influence of UV light.

In order to evaluate which compounds are more favorable to form, the standard ΔH and ΔG of mono- and dichlorination reactions (Fig. 6) and negative logarithm of the equilibrium constant, pK, were analyzed (Table 3). Based on the approach presented by Li *et al.* [47], all of the available ring positions in 4-MCA, 4-MBA, and 4-MP were examined. It is well known that ΔG is a much better criterion of thermodynamic stability because it includes the entropy factor. Generally, the entropy contribution to the ΔG (-TΔS) is the most significant when R=OH or Me and almost negligible when R=Ph (Table 3). Nevertheless, a good correlation coefficient between the calculated values of ΔG and ΔH (r=0.99) indicates that both functions are equivalent in evaluating the thermodynamic stability of chloroorganic compounds.

$H_2O_2$/HCl system is mainly a source of HOCl and $Cl_2$. From the thermodynamic point of view HOCl is a more efficient chlorinating agent than $Cl_2$ (Table 3). This is a general rule since the gas phase enthalpies of monochlorination of benzene with $Cl_2$ and HOCl, calculated from the experimental data [48-51], are -28.85 kcal mol$^{-1}$ and -46.79 kcal mol$^{-1}$ respectively. As one can see from Table 3, reactions of 4-MCA, 4-MBA, and 4-MP with $Cl_2$ and HOCl are in all cases exothermic (ΔH<0) and favorable (ΔG<0). However, the ΔH and ΔG values calculated for the isodesmic reactions (Fig. 6; R=Me, Ph) indicate that chlorination of 4-MCA, 4-MBA, and 4-MP is less favorable than chlorination of methane or benzene.

Mono- and dichlorinated 4-MCA, 4-MBA, and 4-MP derivatives can be classified into six categories: monochloro-4-methoxycinnamic acids (2-Cl-4-MCA and 3-Cl-4-MCA), dichloro-4-methoxycinnamic acids (2,3-diCl-4-MCA, 2,5-diCl-4-MCA, 2,6-diCl-4-MCA, 3,5-diCl-4-MCA), monochloro-4-methoxybenzenealdehydes (2-Cl-4-MBA, 3-Cl-4-MBA),

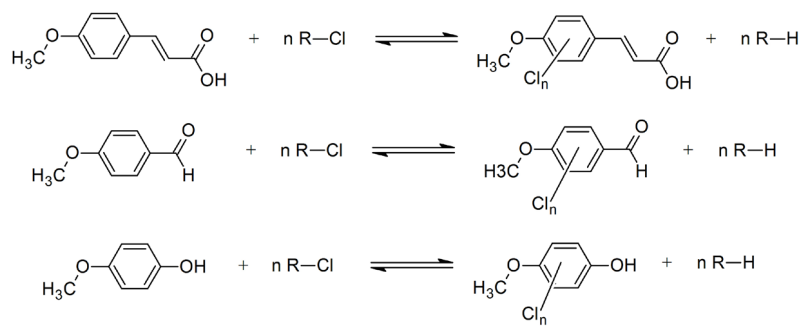

**Figure 6.** Theoretical reactions proposed for thermodynamic stability estimation of chloro-substituted compounds, R=Cl, OH, Me, Ph and n = 1, 2.

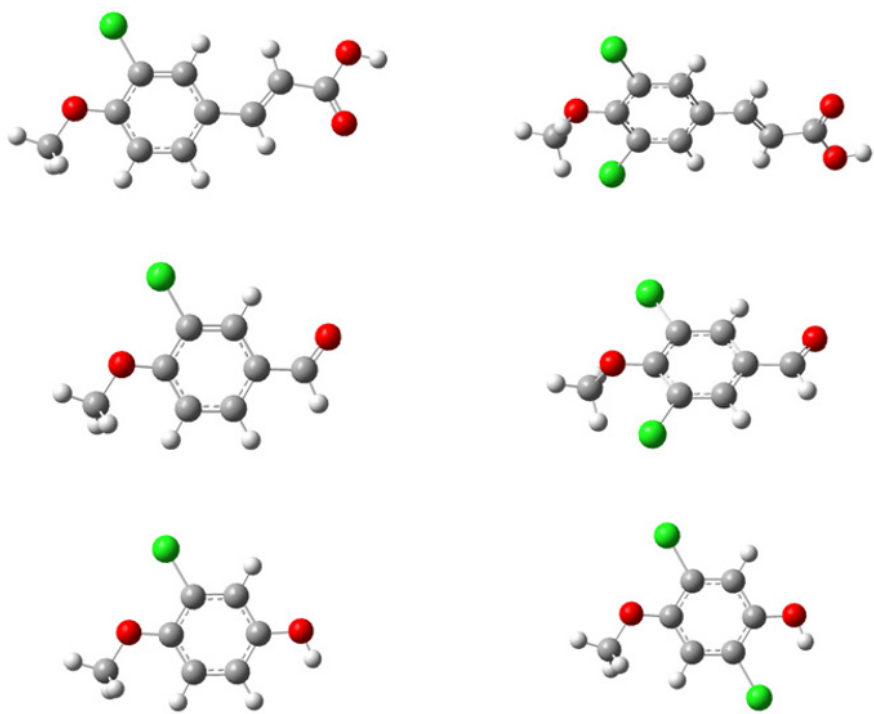

**Figure 7.** Visual representation of mono- and dichlorinated 4-MCA, 4-MBA and 4-MP derivatives detected in the EHMC/$H_2O_2$/HCl/UV system.

dichloro-4-methoxybenzenealdehydes (2,3-diCl-4-MBA, 2,5-diCl-4-MBA, 2,6-diCl-4-MBA, 3,5-diCl-4-MBA), monochloro-4-methoxyphenols (2-Cl-4-MP, 3-Cl-4-MP), and dichloro-4-methoxyphenols (2,3-diCl-4-MP, 2,5-diCl-4-MP, 2,6-diCl-4-MP, 3,5-diCl-4-MP). When comparing two isomers, lower values of ∆H and ∆G indicate higher stability. As we found, chlorination reactions do not in all cases lead to the most thermodynamically favorable product. The most stable isomers were not detected among monochlorinated 4-MP, dichlorinated 4-MCA and 4-MBA derivatives. It is worth to note that in case of 4-MCA and 4-MBA the second chlorination occurs at sterically hindered sites (Fig. 7). Obviously, these pathways seem to be less thermodynamically favorable, which was confirmed by our calculations.

In our experiments the vicinal dichlorinated compounds were not detected in the reaction mixture. Obviously chlorination of 4-MCA and 4-MBA occurs according to the electron-donating resonance effect of the MeO substituent (+R) and the electron-withdrawing resonance effects of the CH=CHCOOH and CHO groups (-R). From the thermodynamic point of view formation of vicinal dichlorinated isomers, 2,3-diCl-4-MCA, 2,3-diCl-4-MBA and 2,3-diCl-4-MP is also less favorable than formation of 2,5-diCl-4-MP, 3,5-diCl-4-MBA and 3,5-diCl-4-MCA detected in the reaction mixture (Table 3). It is worth mentioning that DFT thermochemical calculations presented by Li *et al*. [47] demonstrated that the repulsion of the nearest chlorine atoms in a molecule is a destabilizing factor.

Table 3. Thermodynamic data calculated for monochlorination and dichlorination reactions shown in Fig. 6.

| Product | R | ΔH (kcal mol$^{-1}$) | ΔG (kcal mol$^{-1}$) | pK | Product | R | ΔH (kcal mol$^{-1}$) | ΔG (kcal mol$^{-1}$) | pK |
|---|---|---|---|---|---|---|---|---|---|
| 2-Cl-4-MCA | Cl | -30.15 | -29.69 | -21.78 | 2,5-diCl-4-MBA | Cl | -58.86 | -57.85 | -42.44 |
| | OH | -45.05 | -43.80 | -32.13 | | OH | -88.65 | -86.06 | -63.13 |
| | Me | 1.43 | 2.73 | 2.00 | | Me | 4.31 | 7.00 | 5.13 |
| | Ph | 3.17 | 3.17 | 2.33 | | Ph | 7.78 | 7.88 | 5.78 |
| 3-Cl-4-MCA | Cl | -30.67 | -29.98 | -21.99 | 2,6-diCl-4-MBA | Cl | -50.59 | -49.90 | -36.60 |
| | OH | -45.57 | -44.08 | -32.33 | | OH | -80.38 | -78.11 | -57.30 |
| | Me | 0.91 | 2.45 | 1.80 | | Me | 12.58 | 14.95 | 10.97 |
| | Ph | 2.65 | 2.89 | 2.12 | | Ph | 16.04 | 15.83 | 11.61 |
| 2,3-diCl-4-MCA | Cl | -57.17 | -56.04 | -41.11 | 3,5-diCl-4-MBA | Cl | -55.26 | -54.95 | -40.31 |
| | OH | -86.96 | -84.25 | -61.8 | | OH | -85.06 | -83.16 | -61.00 |
| | Me | 6.00 | 8.81 | 6.46 | | Me | 7.90 | 9.90 | 7.26 |
| | Ph | 9.47 | 9.69 | 7.11 | | Ph | 11.37 | 10.78 | 7.91 |
| 2,5-diCl-4-MCA | Cl | -60.16 | -58.99 | -43.27 | 2-Cl-4-MP | Cl | -31.66 | -31.29 | -22.95 |
| | OH | -89.95 | -87.20 | -63.96 | | OH | -46.55 | -45.39 | -33.30 |
| | Me | 3.00 | 5.85 | 4.29 | | Me | -0.08 | 1.14 | 0.84 |
| | Ph | 6.47 | 6.73 | 4.94 | | Ph | 1.66 | 1.58 | 1.16 |
| 2,6-diCl-4-MCA | Cl | -56.63 | -55.55 | -40.75 | 3-Cl-4-MP | Cl | -31.06 | -30.66 | -22.49 |
| | OH | -86.42 | -83.76 | -61.44 | | OH | -45.95 | -44.77 | -32.84 |
| | Me | 6.54 | 9.30 | 6.82 | | Me | 0.53 | 1.76 | 1.29 |
| | Ph | 10.00 | 10.18 | 7.47 | | Ph | 2.26 | 2.20 | 1.61 |
| 3,5-diCl-4-MCA | Cl | -50.10 | -49.66 | -36.43 | 2,3-diCl-4-MP | Cl | -59.47 | -58.53 | -42.93 |
| | OH | -79.89 | -77.87 | -57.12 | | OH | -89.26 | -86.74 | -63.63 |
| | Me | 13.07 | 15.18 | 11.14 | | Me | 3.70 | 6.32 | 4.64 |
| | Ph | 16.53 | 16.06 | 11.78 | | Ph | 7.17 | 7.20 | 5.28 |
| 2-Cl-4-MBA | Cl | -29.29 | -28.81 | -21.13 | 2,5-diCl-4-MP | Cl | -61.94 | -61.00 | -44.75 |
| | OH | -44.19 | -42.92 | -31.48 | | OH | -91.73 | -89.21 | -65.44 |
| | Me | 2.29 | 3.61 | 2.65 | | Me | 1.23 | 3.84 | 2.82 |
| | Ph | 4.03 | 4.05 | 2.97 | | Ph | 4.70 | 4.72 | 3.46 |
| 3-Cl-4-MBA | Cl | -30.25 | -29.72 | -21.80 | 2,6-diCl-4-MP | Cl | -61.79 | -60.94 | -44.70 |
| | OH | -45.14 | -43.83 | -32.15 | | OH | -91.59 | -89.15 | -65.39 |
| | Me | 1.34 | 2.70 | 1.98 | | Me | 1.37 | 3.91 | 2.87 |
| | Ph | 3.07 | 3.14 | 2.30 | | Ph | 4.84 | 4.79 | 3.51 |
| 2,3-diCl-4-MBA | Cl | -56.04 | -54.90 | -40.27 | 3,5-diCl-4-MP | Cl | -59.24 | -58.65 | -43.02 |
| | OH | -85.83 | -83.11 | -60.96 | | OH | -89.03 | -86.86 | -63.71 |
| | Me | 7.13 | 9.95 | 7.30 | | Me | 3.94 | 6.02 | 4.42 |
| | Ph | 10.60 | 10.83 | 7.94 | | Ph | 7.40 | 7.08 | 5.19 |

### 4.3. Reactivity indices and charge density analysis

It is well known that chemical species which are highly reactive are usually very unstable. Because of this fact, stability of compounds can be measured in terms of reactivity indices [52]. It is important to note that aromatic systems are generally more stable than non-aromatic ones and hence aromaticity indices are often used in evaluation of reactivity and stability [40,53]. Generally, chlorinated 4-MCA, 4-MBA and 4-MP isomers which are more aromatic and which exhibit higher hardness than others were formed under the experimental conditions (Table 4). However, in case of monochlorinated 4-MP derivatives, 3-Cl-4-MP, not 2-Cl-4-MP, was detected (Fig. 4) although 2-Cl-4-MP is slightly more aromatic.

**Table 4.** Reactivity indices derived from HOMO and LUMO energy levels and benzene ring geometry.

| Compound | HOMA | χ [kcal mol⁻¹] | η [kcal mol⁻¹] | ω [kcal mol⁻¹] |
|---|---|---|---|---|
| EHMC | 0.922 | 75.30 | 112.95 | 25.10 |
| 4-MCA | 0.923 | 77.50 | 111.38 | 26.96 |
| 4-MBA | 0.920 | 80.32 | 123.62 | 26.09 |
| 4-MP | 0.966 | 68.71 | 119.54 | 19.75 |
| 2-Cl-4-MCA | 0.917 | 81.26 | 111.38 | 29.64 |
| 3-Cl-4-MCA | 0.919 | 79.69 | 112.32 | 28.27 |
| 2,3-diCl-4-MCA | 0.906 | 82.83 | 111.7 | 30.71 |
| 2,5-diCl-4-MCA | 0.913 | 83.14 | 112.01 | 30.86 |
| 2,6-diCl-4-MCA | 0.911 | 87.22 | 115.46 | 32.95 |
| 3,5-diCl-4-MCA | 0.940 | 82.83 | 116.09 | 29.55 |
| 2-Cl-4-MBA | 0.916 | 85.03 | 122.68 | 29.47 |
| 3-Cl-4-MBA | 0.917 | 83.14 | 123.93 | 27.89 |
| 2,3-diCl-4-MBA | 0.910 | 87.22 | 122.36 | 31.09 |
| 2,5-diCl-4-MBA | 0.915 | 87.22 | 122.36 | 31.09 |
| 2,6-diCl-4-MBA | 0.895 | 88.48 | 124.25 | 31.50 |
| 3,5-diCl-4-MBA | 0.945 | 92.56 | 126.44 | 33.88 |
| 2-Cl-4-MP | 0.966 | 73.10 | 118.91 | 22.47 |
| 3-Cl-4-MP | 0.964 | 72.48 | 119.54 | 21.97 |
| 2,3-diCl-4-MP | 0.955 | 75.61 | 118.91 | 24.04 |
| 2,5-diCl-4-MP | 0.964 | 74.99 | 120.17 | 23.40 |
| 2,6-diCl-4-MP | 0.965 | 77.18 | 118.60 | 25.12 |
| 3,5-diCl-4-MP | 0.966 | 79.38 | 123.93 | 25.42 |

**Table 5.** NBO charges and Fukui function $f_j^-$ values in EHMC, 4-MCA, 4-MBA, 4-MP, 3-Cl-4-MCA, 3-Cl-4-MBA and 3-Cl-4-MP.

| Compound | Position | NBO Charges | $f_j^-$ |
|---|---|---|---|
| 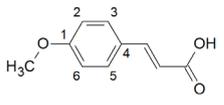 | 1 | 0.341 | 0.13 |
| | 2 | -0.294 | 0.10 |
| | 3 | -0.191 | 0.00 |
| | 4 | -0.144 | 0.22 |
| | 5 | -0.184 | 0.03 |
| | 6 | -0.324 | 0.06 |
| 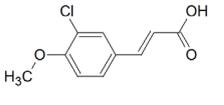 | 1 | 0.334 | 0.13 |
| | 2 | -0.095 | 0.08 |
| | 3 | -0.202 | 0.00 |
| | 4 | -0.126 | 0.20 |
| | 5 | -0.182 | 0.05 |
| | 6 | -0.314 | 0.06 |
| 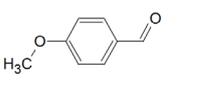 | 1 | 0.359 | 0.12 |
| | 2 | -0.290 | 0.14 |
| | 3 | -0.180 | 0.00 |
| | 4 | -0.223 | 0.27 |
| | 5 | -0.167 | 0.00 |
| | 6 | -0.335 | 0.14 |
| 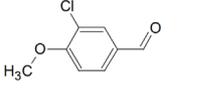 | 1 | 0.334 | 0.12 |
| | 2 | -0.095 | 0.16 |
| | 3 | -0.202 | 0.00 |
| | 4 | -0.126 | 0.23 |
| | 5 | -0.128 | 0.02 |
| | 6 | -0.314 | 0.07 |
| 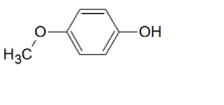 | 1 | 0.281 | 0.17 |
| | 2 | -0.278 | 0.08 |
| | 3 | -0.303 | 0.03 |
| | 4 | 0.273 | 0.19 |
| | 5 | -0.290 | 0.08 |
| | 6 | -0.314 | 0.06 |
| 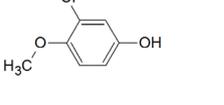 | 1 | 0.276 | 0.17 |
| | 2 | -0.084 | 0.07 |
| | 3 | -0.297 | 0.02 |
| | 4 | 0.290 | 0.17 |
| | 5 | -0.300 | 0.10 |
| | 6 | -0.296 | 0.04 |

As might be expected, the introduction of a chlorine atom into the molecule results in an increase of electronegativity. It means that monochlorinated 4-MCA, 4-MBA and 4-MP are probably less nucleophilic than non-chlorinated ones. Generally speaking the second chlorination step is often kinetically less favorable than the first step. For instance, electrophilic chlorination of benzene using the $FeCl_3/PhNO_2$ system is several times faster than in the case of chlorobenzene [54]. According to the Hammond postulate [55] reaction intermediates are energetically close to the transition states, which means that the formation of stable σ-complexes is kinetically advantageous. Obviously, the charge density on carbon atoms determines the stability of the σ-complexes with electrophiles. One of the useful methods for evaluating susceptibility of certain atoms in a molecule to either nucleophilic or electrophilic attack is a charge distribution analysis [56-60]. Our calculations demonstrated that chlorination occurs at the atoms with the highest electron density (Table 5). This is consistent with the resonance effects of the substituents attached to the benzene ring. In case of 4-MCA and 4-MBA the electron-donating methoxyl group (MeO) is conjugated with the electron-withdrawing substituent, CH=CHCOOH and CHO respectively. The substituent constants $\sigma_p^-$ of MeO, CH=CHCOOH, and CHO groups are -0.26, 0.62, and 1.03 respectively [61]. This shows that C-2 and C-6 atoms (numeration according to Table 5) in 4-MCA and 4-MBA are the most susceptible to substitution by the chlorinating agent. From the local reactivity point of view nucleophilicity of particular atoms in the molecule can be evaluated by means of the Fukui function $f_j^-$. As we found, chlorination reactions occur at the atoms with the highest values of $f_j^-$ among all sites available for substitution. As it can be expected, the electrophilic attack of a chlorinating agent is, in case of 3-Cl-4-MP, the most probable at C-5 atom. This is of course determined by the electron-donating resonance effects (+R) of the OH and Cl groups attached to the benzene ring ($\sigma_R$ of OH and Cl substituents are -0.43 and -0.16 respectively [61]). Besides, the calculated $f_j^-$ values also indicate that C-5 atom is more nucleophilic than C-3 or C-6.

Experimental data demonstrated that EHMC photostability decreases in the presence of ROCS. Spectrophotometric measurements indicated that the reaction proceeds very slowly without UV irradiation. This shows that EHMC is relatively stable in the presence of ROCS when photolysis efficiency is reduced.

The mechanism of 4-MCA, 4-MBA, and 4-MP chlorination is probably complex and consists of radical and ionic steps, including photoinduced C–O bond cleavage, oxidation, and chlorination. DFT thermochemical calculations demonstrated that reactions of 4-MCA, 4-MBA, and 4-MP with $Cl_2$ and HOCl are in all cases thermodynamically probable ($\Delta G<0$) and exothermic ($\Delta H<0$). However, the thermodynamic analysis does not provide a sufficient explanation of chloroorganic compounds formation. In most cases chlorinated 4-MCA, 4-MBA, and 4-MP isomers, which are "harder" than others, were identified in the reaction mixture. This is consistent with the "Maximum Hardness Principle", according to which stable chemical species are less polarizable than unstable ones [52]. In order to better understand the mechanism of formation of chloroorganic compounds the local nucleophilicity was described by means of the Fukui function.

## 5. Conclusions

In this study we examined the effect of common oxidizing and chlorinating agents on EHMC photodegradation.

## Acknowledgements

We thank the Academic Computer Center in Gdańsk for use of their computational facilities.

### References

[1] Y. Dobashi, T. Yuyama, Y. Ohkatsu, Polym. Degrad. Stab. 92, 1227 (2007)
[2] V.S. Sivokhin, Polym. Sci. USSR 21, 1207 (1980)
[3] A. Kikuchi, H. Saito, M. Mori, M. Yagi, Photochem. Photobiol. Sci. 10, 1902 (2011)
[4] A.J.M. Santos, M.S. Miranda, J.C.G. Esteves da Silva, Water Res. 46, 3167 (2012)
[5] A.J.M. Santos, D.M.A. Crista, M.S. Miranda, I.F. Almeida, J.P. Sousa e Silva, P.C. Costa, M.H. Amaral, P.A.L. Lobão, J.M. Sousa Lobo, J.C.G. Esteves da Silva, Environ. Chem. 10, 127 (2013)
[6] M.S. Miranda, L. Pinto da Silva, J.C.G. Esteves da Silva, J. Phys. Org. Chem. 27, 47 (2014)
[7] Y.T. Chen, Y.T. Lin, C.C. Li, S.F. Sie, Y.W. Chen-Yang, Colloids Surf. B 115, 191 (2014)
[8] Z.A. Lewicka, W.W. Yu, B.L. Oliva, E.Q. Contreras, V.L. Colvin, J. Photochem. Photobiol. A, 263, 24 (2013)
[9] L. Zhou, Y. Ji, C. Zeng, Y. Zhang, Z. Wang, X. Yang, Water Res., 47, 153 (2013)
[10] Y. Ji, L. Zhou, Y. Zhang, C. Ferronato, M. Brigante, G. Mailhot, X. Yang, J.-M. Chovelon, Water Res. 47, 5865 (2013)
[11] A. Christiansson, J. Eriksson, D. Teclechiel, Å. Bergman, Environ. Sci. Pollut. Res. 16, 312 (2009)
[12] L.A. MacManus-Spencer, M.L. Tse, J.L. Klein, A.E. Kracunas, Environ. Sci. Technol. 45, 3931 (2011)
[13] C. Ferrari, H. Chen, R. Lavezza, C. Santinelli, I. Longo, E. Bramanti, Int. J. Photoenergy 2013, 1 (2013)
[14] V. Nikolić, D. Ilić, L. Nikolić, M. Stanković, M. Cakić, L. Stanojević, A. Kapor, M. Popsavin, Cent. Eur. J. Chem. 8, 744 (2010)
[15] A.Y.C. Tong, R. Braund, D.S. Warren, B.M. Peake,

Cent. Eur. J. Chem. 10, 989 (2012)
[16] S. Pattanaargson, P. Limphong, Int. J. Cosmetic. Sci. 23, 153 (2001)
[17] N. Tarras-Wahlberg, G. Stenhagen, O. Larkö, A. Rosén, A.-M. Wennberg, O. Wennerström, J. Invest. Dermatol. 113, 547 (1999)
[18] J. Gaca, S. Żak Hydrogen peroxide and chlorides, examples of application and theoretical aspects (University of Technology and Agriculture in Bydgoszcz Publishers, Poland, 2004) (in Polish)
[19] M. Nakajima, T. Kawakami, T. Niino, Y. Takahashi, S. Onodera, J. Health. Sci. 55, 363 (2009)
[20] A. Gackowska, J. Gaca, Chemik 4, 301 (2011)
[21] N. Higashi, A. Ikehata, N. Kariyama, Y. Ozaki, Appl. Spectrosc. 62, 1022 (2008)
[22] P. Cysewski, A. Gackowska, J. Gaca, Chemosphere 63, 165 (2006)
[23] C.W. Jones, Applications of hydrogen peroxide and derivatives (Royal Society of Chemistry, Cambridge 1999)
[24] A.D. Becke, Phys. Rev. A 38, 3098 (1988)
[25] A.D. Becke, J. Chem. Phys. 98, 5648 (1993)
[26] B. Miehlich, A. Savin, H. Stoll, H. Preuss, Chem. Phys. Lett. 157, 200 (1989)
[27] C. Lee, W. Yang, R.G. Parr, Phys. Rev. B 37, 785 (1988)
[28] S. Miertuš, E. Scrocco, J. Tomasi, Chem. Phys. 55, 117 (1981)
[29] S. Miertuš, J. Tomasi, Chem. Phys. 65, 239 (1982)
[30] M.J. Frisch, G.W. Trucks, H.B. Schlegel, G.E. Scuseria, M.A. Robb, J.R. Cheeseman, J.A. Montgomery, Jr., T. Vreven, K.N. Kudin, J.C. Burant, J.M. Millam, S.S. Iyengar, J. Tomasi, V. Barone, B. Mennucci, M. Cossi, G. Scalmani, N. Rega, G.A. Petersson, H. Nakatsuji, M. Hada, M. Ehara, K. Toyota, R. Fukuda, J. Hasegawa, M. Ishida, T. Nakajima, Y. Honda, O. Kitao, H. Nakai, M. Klene, X. Li, J.E. Knox, H.P. Hratchian, J.B. Cross, C. Adamo, J. Jaramillo, R. Gomperts, R.E. Stratmann, O. Yazyev, A.J. Austin, R. Cammi, C. Pomelli, J.W. Ochterski, P.Y. Ayala, K. Morokuma, G.A. Voth, P. Salvador, J.J. Dannenberg, V.G. Zakrzewski, S. Dapprich, A.D. Daniels, M.C. Strain, O. Farkas, D.K. Malick, A.D. Rabuck, K. Raghavachari, J.B. Foresman, J.V. Ortiz, Q. Cui, A.G. Baboul, S. Clifford, J. Cioslowski, B.B. Stefanov, G. Liu, A. Liashenko, P. Piskorz, I. Komaromi, R.L. Martin, D.J. Fox, T. Keith, M.A. Al-Laham, C.Y. Peng, A. Nanayakkara, M. Challacombe, P.M.W. Gill, B. Johnson, W. Chen, M.W. Wong, C. Gonzalez, J.A. Pople, Gaussian 03, Revision E. 01 (Gaussian, Inc., Pittsburgh PA, 2004)
[31] R.G. Pearson, Proc. Natl. Acad. Sci. USA 83, 8440 (1986)
[32] R.G. Parr, L. von Szentpály, S. Liu, J. Am. Chem. Soc. 121, 1922 (1999)
[33] C. Møller, M.S. Plesset, Phys. Rev. 46, 0618 (1934)
[34] M. Head-Gordon, J.A. Pople, M.J. Frisch, Chem. Phys. Lett. 153, 503 (1988)
[35] S. Saebø, J. Almlöf, Chem. Phys. Lett. 154, 83 (1989)
[36] M.J. Frisch, M. Head-Gordon, J.A. Pople, Chem. Phys. Lett. 166, 275 (1990)
[37] M.J. Frisch, M. Head-Gordon, J.A. Pople, Chem. Phys. Lett. 166, 281 (1990)
[38] M. Head-Gordon, T. Head-Gordon, Chem. Phys. Lett. 220, 122 (1994)
[39] J. Kruszewski, T.M. Krygowski, Tetrahedron Lett. 13, 3839 (1972)
[40] T.M. Krygowski, M.K. Cyrański, Z. Czarnocki, G. Haäfelinger, A.R. Katritzky, Tetrahedron 56, 1783 (2000)
[41] W. Yang, W.J. Mortier, J. Am. Chem. Soc. 108, 5708 (1986)
[42] S. Pattanaargson, T. Munhapol, P. Hirumsupachot, P. Luangthongaram, J. Photochem. Photobiol. A 161, 269 (2004)
[43] S.P. Huong, V. Andrieu, J.-P. Reynier, E. Rocher, J.-D. Fourneron, J. Photochem. Photobiol. A 186, 65 (2007)
[44] P. Renard, F. Siekmann, A. Gandolfo, J. Socorro, G. Salque, S. Ravier, E. Quivet, J.-L. Clément, M. Traikia, A.-M. Delort, D. Voisin, V. Vuitton, R. Thissen, A. Monod, Atmos. Chem. Phys. 13, 6473 (2013)
[45] N. De la Cruz, J. Giménez, S. Esplugas, D. Grandjean, L.F. de Alencastro, C. Pulgarín Water Res. 46, 1947 (2012)
[46] M. Czaplicka, J. Hazard. Mater. 134, 45 (2006)
[47] X.-W. Li, E. Shibata, T. Nakamura, Mater. Trans. 44, 2441 (2003)
[48] M.V. Roux, M. Temprado, J.S. Chickos, Y. Nagano, J. Phys. Chem. Ref. Data 37, 1855 (2008)
[49] V.A. Platonov, Yu.N. Simulin Russ. J. Phys. Chem. 59, 179 (1985)
[50] J.D. Cox, D.D. Wagman, V.A. Medvedev, CODATA Key Values for Thermodynamics (Hemisphere Publishing Corp, New York, 1984)
[51] M.W. Chase, NIST-JANAF Thermochemical Tables, J. Phys. Chem. Ref. Data, Monograph 9, 4th edition (American Chemical Society, American Institute of Physics, Washington, DC, 1998)
[52] M. Bekbolet, Z. Çınar, M. Kılıç, C.S. Uyguner, C. Minero, E. Pelizzetti, Chemosphere 75, 1008 (2009)

[53] E. Zahedi, S. Ali-Asgari, V. Keley, Cent. Eur. J. Chem. 8, 1097 (2010)
[54] R. Taylor, in: C.H. Bamford, C.F.H. Tipper (Eds.), Comprehensive chemical kinetics, (Elsevier Publishing Co, Amsterdam/New York, 1972) Volume 13: Reactions of aromatic compounds, 1
[55] G.S. Hammond, J. Am. Chem. Soc. 77, 334 (1955)
[56] T.E. Jones, J. Phys. Chem. A 116, 4233 (2012)
[57] Md.E.U. Hoque, H.W. Lee, Bull. Korean. Chem. Soc. 32, 2109 (2011)
[58] M. Przybyłek, J. Gaca, Chem. Pap. 66, 699 (2012)
[59] V. Benin, J. Mol. Struc. (Theochem) 764, 21 (2006)
[60] H.D. Banks, J. Org. Chem. 68, 2639 (2003)
[61] C. Hansch, A. Leo, R.W. Taft, Chem. Rev. 91, 165 (1991)